\documentclass[12pt]{article}

\usepackage{latexsym}

\font\tenmsbm=msbm10 scaled 1200
\font\sevenmsbm=msbm9
\newfam\msbmfam
\textfont\msbmfam=\tenmsbm
\scriptfont\msbmfam=\sevenmsbm
\def\msbm{\fam\msbmfam\tenmsbm}

\makeatletter
\@addtoreset{equation}{section}
\makeatother

\newcommand{\eref}[1]{(\ref{#1})}

\def\beq{\begin{equation}}
\def\eeq{\end{equation}}
\def\bea{\begin{eqnarray}}
\def\eea{\end{eqnarray}}
\def\bet{\begin{tabular}}
\def\eet{\end{tabular}}

\def\pa{{\partial}}

\catcode`@=11
\def\lsim{\mathchoice
  {\mathrel{\lower.8ex\hbox{$\displaystyle\buildrel<\over\sim$}}}
  {\mathrel{\lower.8ex\hbox{$\textstyle\buildrel<\over\sim$}}}
  {\mathrel{\lower.8ex\hbox{$\scriptstyle\buildrel<\over\sim$}}}
  {\mathrel{\lower.8ex\hbox{$\scriptscriptstyle\buildrel<\over\sim$}}} }
\def\gsim{\mathchoice
  {\mathrel{\lower.8ex\hbox{$\displaystyle\buildrel>\over\sim$}}}
  {\mathrel{\lower.8ex\hbox{$\textstyle\buildrel>\over\sim$}}}
  {\mathrel{\lower.8ex\hbox{$\scriptstyle\buildrel>\over\sim$}}}
  {\mathrel{\lower.8ex\hbox{$\scriptscriptstyle\buildrel>\over\sim$}}} }
\def\croce{\displaystyle / \kern-0.2truecm\hbox{$\backslash$}}
\def\lqua{\lower4pt\hbox{\kern5pt\hbox{$\sim$}}\raise1pt
\hbox{\kern-8pt\hbox{$<$}}~}
\def\gqua{\lower4pt\hbox{\kern5pt\hbox{$\sim$}}\raise1pt
\hbox{\kern-8pt\hbox{$>$}}~}
\def\mma{\lower1pt\hbox{\kern5pt\hbox{$\scriptstyle <$}}\raise2pt
\hbox{\kern-7pt\hbox{$\scriptstyle >$}}~}
\def\mmb{\lower1pt\hbox{\kern5pt\hbox{$\scriptstyle >$}}\raise2pt
\hbox{\kern-7pt\hbox{$\scriptstyle <$}}~}
\def\mmc{\lower4pt\hbox{\kern5pt\hbox{$<$}}\raise1pt
\hbox{\kern-8pt\hbox{$>$}}~}
\def\mmd{\lower4pt\hbox{\kern5pt\hbox{$>$}}\raise1pt
\hbox{\kern-8pt\hbox{$<$}}~}
\def\lsu{\raise4pt\hbox{\kern5pt\hbox{$\sim$}}\lower1pt
\hbox{\kern-8pt\hbox{$<$}}~}
\def\gsu{\raise4pt\hbox{\kern5pt\hbox{$\sim$}}\lower1pt
\hbox{\kern-8pt\hbox{$>$}}~}
\def\croce{\displaystyle / \kern-0.2truecm\hbox{$\backslash$}}
\def\ali{\hbox{A \kern-.9em\raise1.7ex\hbox{$\scriptstyle \circ$}}}
\def\2frecce{\hbox{\lower 0.3ex\hbox{$\leftarrow$} 
\hbox{\kern-1.3em\raise 0.3ex\hbox{$\rightarrow$}}}}
\def\quad@rato#1#2{{\vcenter{\vbox{
        \hrule height#2pt
        \hbox{\vrule width#2pt height#1pt \kern#1pt \vrule width#2pt}
        \hrule height#2pt} }}}
\def\quadratello{\mathchoice
\quad@rato5{.5}\quad@rato5{.5}\quad@rato{3.5}{.35}\quad@rato{2.5}{.25} }
\def\bbox{\quadratello}

\begin{document}

\begin{titlepage}

\begin{flushright}
Preprint DFPD 98/TH/37\\
hep-th/9808025\\
August 1998\\
\end{flushright}

\vspace{2truecm}

\begin{center}

{\Large \bf Self--dual tensors and gravitational anomalies 
in $4n+2$ dimensions}

\vspace{1cm}

{Kurt Lechner\footnote{kurt.lechner@pd.infn.it}}

\vspace{1cm}

\medskip

{ \it Dipartimento di Fisica, Universit\`a degli Studi di 
Padova,

\smallskip

and

\smallskip

Istituto Nazionale di Fisica Nucleare, Sezione di Padova, 

Via F. Marzolo, 8, 35131 Padova, Italia}

\vspace{1cm}

\begin{abstract}

Starting from a manifestly Lorentz-- and diffeomorphism--invariant 
classical action we perform a perturbative derivation
of the gravitational anomalies for chiral bosons 
in $4n+2$ dimensions. The manifest classical invariance is achieved using
a newly developed method based on a scalar auxiliary field and two
new bosonic local symmetries. The resulting anomalies coincide with the
ones predicted by the index theorem.  
In the two--dimensional case, moreover, we perform an exact 
covariant computation
of the effective action for a chiral boson (a scalar) which is seen to 
coincide with the effective action for a two--dimensional complex 
Weyl--fermion. All these results support the quantum reliability of the
new, at the classical level manifestly invariant, method.

\end{abstract}

\end{center}
\end{titlepage}

\newpage

\baselineskip 6 mm


\section{Introduction}

For long time the main problem related with (anti)self--dual antisymmetric
tensors of rank $2n+1$ in $4n+2$ dimensions (chiral bosons),
was the absence of a manifestly Lorentz--invariant action principle.
Some time ago for two--dimensional chiral bosons Siegel \cite{Siegel}
proposed a manifestly invariant lagrangian which is, however, plagued 
by a quantum anomaly whose proper handling constitutes still a problematic
feature of his approach. Apart from this feature, the most inconvenient 
aspect of this method, also in higher dimensions \cite{armeni},  
is that it produces the square of the self--duality equation of motion 
rather than the self--duality condition itself. This leads at the 
quantum level to problematic aspects due to the presence of second class 
constraints which have to be appropriately handled \cite{HT}.

Manifestly covariant actions for chiral bosons have also been constructed in 
\cite{MWY} using an infinite tower of Lagrange multipliers, in which case
the problem is shifted to a consistent truncation of this tower.

Another class of actions, which avoid all these problems, has been 
presented in
\cite{HT,FJ,BN1,BN2,S}. The principal drawback of these actions is that
they are not manifestly Lorentz--invariant, although being invariant under
a set of modified transformations which satisfy the Lorentz algebra.
When one couples these actions to gravity, due to 
this feature, they lack also manifest invariance  under diffeomorphisms
and a detailed analysis is needed to establish it \cite{HT}.
Clearly this non manifest invariance becomes even more problematic at
the quantum level. Nevertheless, in \cite{BN2,B}, using these actions
the gravitational anomalies for chiral bosons have been derived and shown
to coincide with the expected results \cite{AGW,AGG}.

In this paper we rederive the gravitational anomalies for
chiral bosons in $4n+2$ dimensions using a newly developed lagrangian 
approach \cite{PST} which, at the classical level, is manifestly invariant 
under Lorentz--transformations. This feature makes a manifestly 
diffeomorphism invariant coupling to gravity trivial: the minimal
prescription just works. The method itself is based on a single
scalar auxiliary field $a(x)$ and on two new bosonic symmetries whose 
physical interpretation is very simple: the first symmetry allows to 
eliminate the auxiliary field and the second reduces the second order
equation of motion for the antisymmetric tensor to the first order
self--duality condition. 

At the classical level this new method appears to be very general. It
proved compatible with all known symmetries, e.g. with global \cite{KL}
and local supersymmetry \cite{DLT}, with $\kappa$--symmetry \cite{M5}
and with manifest duality symmetry between $p$--form potentials in a generic
space--time dimension $D$ and their Hodge duals, 
$(D-p-2)$--forms \cite{DLT,Maxwell,D11}; it is also consistent with
(self)interacting chiral bosons \cite{PST1,IIB} and, as we will see in 
detail in section two, the two bosonic symmetries  allow a simple control 
of the physical propagating degrees of freedom of the chiral bosons.
Moreover, if one chooses a gauge fixing for the scalar field such that 
$a(x)=n_m x^m$, where $n_m$ is a constant vector, one recovers the 
non--manifestly covariant formulations mentioned above: 
$n_m=(1,0,\ldots,0)$ leads to \cite{HT,FJ,BN1,BN2} and 
$n_m=(0,0,\ldots,1)$ leads to \cite{S}.

Due to these successes of the method at the classical level the most
compelling question which remains is if it has some reliability also 
at the quantum level. For dimensions greater or equal than six,
of course, 
this question is somehow academic due to non--renormalizability of
the actions; in two dimensions, on the other side, the question is
perfectly well suited and the expected results are known: e.g. the 
effective action for a chiral boson coupled to gravity should equal  
the effective action of a complex Weyl fermion. Moreover, the derivation
of gravitational anomalies is a well suited issue also in higher 
dimensions, since anomalies are always finite,
and the expected results for chiral bosons are known \cite{AGW,AGG}. 

The purpose of this paper is to perform a quantum check 
of the new method in these directions. In the next section we present
the classical covariant action for chiral bosons in $D=4n+2$ dimensions, 
interacting 
with an external gravitational field, and show that it gives rise, as 
equation of motion, to the self--duality condition for the field--strength.
In this section we give also a self--contained account of the new covariant
method itself.
To avoid writing indices we will use mainly the language of forms.
Section three is dedicated to a detailed analysis for chiral bosons in 
$D=2$. We perform an exact covariant computation for the effective action 
and show that it equals, modulo local terms, the effective action of a 
complex Weyl fermion, computed in a diffeomorphism preserving framework
(i.e. where the anomaly is shifted from diffeomorphisms to local Lorentz--
and Weyl--transformations). A perturbative one--loop computation of the
$D=2$ anomaly is also performed to prepare the anomaly derivation in
higher dimensions. In section four we derive the gravitational anomalies 
for chiral bosons in higher dimensions showing that the effective Feynman
rules associated to our classical action coincide with the 
Feynman rules conjectured in \cite{AGW} which, in turn, led to the results
predicted by the index theorem \cite{AGG}. This procedure requires, in 
particular, an appropriate gauge fixing of the two new bosonic symmetries
mentioned above. Section five is devoted to some concluding remarks.
\vskip1truecm

\section{The classical action for chiral bosons}

The language we will use mainly in this paper is the language of 
differential forms.
This will allow to keep the formulae compact, i.e. 
without writing explicitly indices,  and to perform 
the relevant computations using only the algebra satisfied by 
the exterior differential $d$, the hodge--dual $*$ and the interior
product $i_v$ of a vector $v$ with a $p$--form (see below). 

In particular we will write our classical action for chiral bosons 
in $D=2k+2$ dimensions, with $k$ even, as an integral over a $D$--form.
Before doing that we state our conventions on forms and present the
relevant identities.

We define the components of a $p$--form $\phi_p$ as
$$ 
\phi_{p} =
\frac{1}{p!} dx^{n_1} \cdots dx^{n_p} \phi_{n_{p} \cdots n_1},
$$
and correspondingly the exterior differential $d=dx^n\pa_n$ begins to act 
on the right. The product between forms will always be the wedge
product and the symbol $\wedge$ will be omitted.

The interior product of a vector field $v=v^n\pa_n$ with a $p$--form 
is defined by
$$
i_v\phi_p =
\frac{1}{(p-1)!} dx^{n_1} \cdots dx^{n_{p-1}}v^{n_p} \phi_{n_{p} 
\cdots n_1},
$$
and satisfies the same distribution law, $i_v(\phi_p \phi_q)=
\phi_p i_v\phi_q +(-)^q i_v(\phi_p) \phi_q$, as $d$.

Introducing a metric $g^{mn}$ on the space our convention for the 
Hodge--dual is 
$$
*(\phi_{p}) \equiv \frac{1}{(D - p)!} dx^{n_1} \cdots dx^{n_{D-p}}
(*\phi)_{n_{D-p} \cdots n_1},
$$
where
$$
(*\phi)_{n_1 \cdots n_{D-p}} = \frac{1}{p!} 
g_{n_1m_1}\cdots g_{n_{D-p}m_{D-p}}
{\varepsilon^{m_1\cdots m_{D-p}j_1\cdots j_p} \over \sqrt{g}}
\phi_{j_1 \cdots j_p}
$$
and $g=-det g_{mn}$. Our flat metric is $\eta_{mn}=(1,-1,\cdots,-1)$
and $\varepsilon^{01\cdots D-1}=+1$. 

The delta--operator, which sends a
$p$--form in a $(p-1)$--form, is defined as usual by
$$
\delta = *d*.
$$
On a $p$--form the square of the Hodge--dual, in an even dimensional 
space--time as the ones considered here, satisfies
\beq
*^2=(-)^{p+1}.
\eeq
 
Using these properties and definitions one can prove the following 
operatorial identities, which hold on any $p$--form and for
any vector field $v$, and which will
be frequently used in what follows:
\bea
\label{v1}
*i_v&=&v*\\
\label{v2}
i_v*&=&-*v\\
\label{v3}
i_vv-vi_v&=&(-)^p v^2\\
\label{v4}
\bbox_g&=&D_m g^{mn}D_n=\delta d +d\delta.
\eea
With the one--form $v$ we mean here 
$$
v=dx^ng_{mn}v^m,
$$
and $v^2=g_{mn}v^mv^n$. Particularly useful will be the following
decompositions of the identity ${\msbm I}$ and of the $*$--operator 
which follow from these identities: 
\bea
\label{dec1}
{\msbm I}&=&{1\over v^2}\left((-)^{p+1}vi_v+*vi_v*\right)\\
\label{dec*}
*&=&(-)^{p+1} {1\over v^2}\left(*vi_v+vi_v*\right)\\
\label{v33}
(1-*)vi_v(1-*)&=&v^2(1+(-)^p*)+vi_v(1+(-1)^p).
\eea
\eref{dec1} allows, in particular, to decompose every $p$--form uniquely
into a component along $v$ and a component orthogonal to $v$.

We turn now to the construction of the action \cite{PST}. 
Chiral bosons are described 
by a $k$--form $B$ whose curvature, a $(k+1)$--form, 
$$
H=dB
$$
satisfies as equation of motion the self--duality condition
\beq\label{self}
H=*H.
\eeq

For definiteness we treat here the case of a self--dual field strength,
for an antiself--dual field strength the procedure is completely analogous.

As anticipated in the introduction, in order to write an action
in addition to $B$ we introduce also
an auxiliary scalar field $a(x)$ and define the related one-form 
$$
v = {da\over \sqrt{g^{mn}\pa_m a\pa_n a}}\equiv dx^nv_n,
$$
which satisfies 
$$
v^2=g_{mn}v^mv^n=1;
$$
this leads, in particular, in \eref{dec1} -- \eref{v33} to 
the disappearance of the factor $v^2$.

The action for a chiral boson in an external gravitational field depends
now also on $a$ and is given by 
\beq
\label{az}
S[B,a,g]={1\over 2}\int v\,h\,H ={1\over 4}\int\left(H*H-h*h\right),
\eeq
where we defined the $k$--form
$$
h=i_v(H-*H).
$$
The equality of the two expressions in this formula can be inferred
from the definition of $h$ and from the decomposition \eref{dec1},
which leads to
\beq
\label{ide}
H-*H=vh-*vh.
\eeq

From the second way of writing $S$ one sees that this action equals
the action for non--chiral bosons, modulo a term which is proportional
to the square of the self--duality condition \eref{self}. This particular
form of the action is dictated by the symmetries it has. 
Under generic variations of $B$ and $a$ we get, in fact
\beq
\label{var}
\delta S=\int\left(v\,h \,d\delta B -{1\over 2}{1\over 
\sqrt{g^{mn}\pa_m a\pa_n a}}\,v\,h\,h\,d\delta a\right).
\eeq
From this formula it is easy to see that the action is invariant
under the following local transformations ($\delta g^{mn}=0$):   
\bea
1) &  &\delta B =  \frac{1}{\sqrt{g^{mn}\pa_m a\pa_n a}}\, h\,\varphi, \qquad 
\delta a = \varphi,\\
2) &  &\delta B = \Lambda_{k-1}\,da,  \qquad \qquad\qquad \delta a = 0,\\
3) &  &\delta B = d\Sigma_{k-1},  \quad\qquad\qquad \qquad \delta a = 0.
\eea
Here $\Sigma_{k-1}$ and $\Lambda_{k-1}$ are $(k-1)$--form transformation
parameters and $\varphi$ is a scalar transformation parameter; under 2) and
3) the action is, actually, also invariant under {\it finite} 
transformations.

The transformation 3) is nothing else than the usual gauge invariance
for $k$--forms, $B$ appears in \eref{az}, in fact, only through its field
strength $dB$. The transformation 1) says that the auxiliary field $a$ is
a non propagating
"pure gauge" field in that it can be transformed to any arbitrary value; 
due to 
its non--polynomial appearance in the action, however, mainly the
appearance of the factor ${g^{mn}\pa_m a\pa_n a}$ at the denominator, 
it can not be set to zero. The transformation 2), instead, allows to 
reduce the equation of motion for $B$ to \eref{self}. To see this
we read the equations of motion for $a$ and $B$ respectively from \eref{var}
\bea
\label{ema}
d \left(\frac{1}{\sqrt{g^{mn}\pa_m a\pa_n a}}\,v\,h\,h  \right) &=& 0, \\
\label{emb}
d (v\,h) &=& 0.
\eea
It can be directly checked that \eref{ema} is a consequence of \eref{emb}, as
follows also from the fact that $a$ is pure gauge. 

The  general solution of \eref{emb}, on the other hand, is given by
\beq
\label{sol}
v\,h=d\tilde \Lambda_{k-1}da
\eeq
for some $(k-1)$--form $\tilde \Lambda_{k-1}$. Since under a finite
transformation 2) we find
$$
v \,h \to v\, h +  d \Lambda_{k-1} \, da,
$$
choosing $\Lambda_{k-1}=\tilde \Lambda_{k-1}$ this transformation can 
be used to reduce \eref{sol} to
$$
h=0.
$$
Due to \eref{ide} this is then equivalent to the self--duality equation
of motion.

This concludes the proof that our action describes correctly the dynamics
of classical chiral bosons 
interacting with an external gravitational field. The
(gauge--fixed) equation of motion we got, $H=*H$, is manifestly 
invariant and $a$--independent and has been obtained using the equations 
of motion and the symmetries of the action. In particular, for the
symmetry 2) we used a rather unconventional "gauge-fixing".
On the other hand, at the quantum level, in a functional integral approach, 
one can not
make direct use of the equations of motion and needs conventional
gauge--fixings, i.e. gauge fixings of the type $f(B,a)=0$. 
In preparation of the quantum developments of sections three and four 
we present here such a set of gauge fixings and show that it leads to the
correct number of physical degrees of freedom carried by chiral bosons 
in $2k+2$ dimensions. 

We choose a flat metric $g^{mn}=\eta^{mn}$ and fix the symmetries
1)--3) according to
\bea
1') & & a(x)= n_ix^i\\
2') & & i_n  B = 0 \qquad \leftrightarrow \qquad n^{i_1}B_{i_1\dots i_k}=0\\
3') & & \delta B = 0 \qquad \leftrightarrow \qquad 
\pa^{i_1}B_{i_1\dots i_k}=0,
\eea
where $n_i$ is a constant vector, normalized such that 
$n^2=n_in_j\eta^{ij}=1$.
The gauge fixing for the symmetry 1) implies, in particular, that 
$v={da\over {\sqrt n^2}}=dx^in_i\equiv n$. The 
choice $1')$ appears the most simplest and treatable one and breaks manifest
Lorentz invariance; manifestly invariant gauge fixings for this symmetry
do not seem to exist.

For what concerns 
$2')$ we observe that \eref{dec1} allows to decompose $B$  as
$$
B= -ni_nB +*\,ni_n*B
$$
and that the symmetry 2) shifts the component along $n$, the first one, 
by an arbitrary amount leaving the second one invariant. The  choice $2')$ 
amounts then just to setting the component along $n$ to zero.
The gauge fixing for 3) is just the usual covariant Lorentz gauge.

The gauge fixings $2')$ and $3')$ leave a "residual" invariance for which
$$
\delta_{res} B=n\Lambda_{k-1}+d\Sigma_{k-1}
$$
with the constraint
$$
i_n(n\Lambda_{k-1}+d\Sigma_{k-1})=0=d*(n\Lambda_{k-1}+d\Sigma_{k-1}).
$$

Using only $1')$
the equation of motion for $B$, \eref{emb}, becomes now
\beq
\label{emb3}
(T\pa_n+T^2)B=0,
\eeq
where $\pa_n= n^i\pa_i$ operates only on the components of a form and
the operator $T$, which sends a $k$--form in a $k$--form, is defined
by
\beq
\label{T}
T=*\,n\,d=*\,d\,n=-i_n*d. 
\eeq
This operator, which on the components of a $k$--form acts as a 
$k\times k$ antisymmetric 
tensor,  will play a central role in section four, so we present
here its main properties. 
Viewed as a tensor, $T$ is symmetric in the interchange of the two branches
of $k$ antisymmetric indices.
Its square, using the algebra given 
above, can be computed to be (on an even form)
\beq
\label{t2}
T^2= (\pa_n^2-\bbox)+(n\pa_n+d)\delta-(n\delta+\pa_n)di_n.
\eeq
Applying $T$ again to this formula, due to $(i_n)^2=d^2=\delta^2 =0$, only
the first bracket contributes and one gets
\beq
\label{t3}
T^3=(\pa_n^2-\bbox)T, 
\eeq
which is the main formula. Also, on forms which satisfy $2')$ and $3')$
we have 
\beq
\label{it2}
T^2 B=(\pa_n^2-\bbox)B,
\eeq
and \eref{emb3} reduces to
\beq
\label{emgf}
(\pa_n^2-\bbox)B =-T\pa_nB.
\eeq
Squaring the operators appearing in this relation on the left and on the
right hand side and using again \eref{it2}
we obtain
\beq
\bbox\,(\pa_n^2-\bbox)B=0.
\eeq
The solution $(\pa_n^2-\bbox)B=0$ would imply $T\pa_nB=0$ and the solutions
of this equation are "pure gauge'', in the sense that they can be eliminated
using the residual invariances given above. We remain therefore with the
equation 
$$
\bbox \,B=0,
$$
which describes massless excitations as expected. In this case
\eref{emgf} reduces to a constraint on the polarizations
\beq
\label{red}
TB=-\pa_n B.
\eeq
Going to momentum space, 
$B_{i_1\cdots i_k}(x)\rightarrow b_{i_1 \cdots i_k}(p)$,
and choosing for example 
$$
n^i=(1,0,\cdots,0),
$$ 
we split our indices
into $i=(0,\alpha)$. Then $2')$ implies that only space--like indices
survive in the polarizations and our solution is characterized by
\bea
p_ip^i&=&0\\
p^{\alpha_1} b_{\alpha_1 \cdots \alpha_k}&=&0\\
b_{\alpha_1\cdots\alpha_k}&=&{1\over k!}\varepsilon_{\alpha_1\cdots\alpha_k
\beta_1\cdots \beta_{k+1}}{p^{\beta_1}\over |\vec p|}
b^{\beta_2\cdots \beta_{k+1}}.
\eea
The third condition is just \eref{red} in momentum space.
One can easily count the number of independent
polarizations which remain undetermined by these equations and the 
result is ${1\over 2}{(2k)!\over (k!)^2}$, which is the correct number.

This was just a check of the appropriateness of the 
the gauge fixings $1')$--$3')$; they will prove to be very convenient also 
at the quantum level as we will see in section four.
\vskip1truecm

\section{Chiral bosons in $D=2$}

In the two--dimensional case it is convenient to work with light--cone 
indices and to introduce light--cone zweibeins $e_{\pm}^i$
to describe the metric
\bea
g^{ij}&=&{1\over 2}\left(e_-^i e_+ ^j + e_+^i e_- ^j\right)\\
\label{det}
{\varepsilon^{ij}\over \sqrt{g}}
&=&{1\over 2}\left(e_-^i e_+ ^j - e_+^i e_- ^j\right).
\eea
All vector indices can then be transformed to local Lorentz indices
through $V_{\pm}=e_\pm^iV_i$. For example $\pa_{\pm}=e_\pm^i\pa_i$, 
$\pa_\pm a=e_\pm^i\pa_ i a$ and so on.

The action \eref{az} can then be rewritten, in two dimensions,  also as
\beq
\label{az2}
S[B,a,e]={1\over 2}\int d^2x \sqrt{g}\left(\pa_+B\pa_-B 
-{\pa_+a \over\pa_-a}\pa_-B\pa_-B\right), 
\eeq
where $B(x)$ is now a scalar field. 
This action is also invariant under local Weyl rescalings of the metric,
as it should.

The self--duality equation becomes in this case simply 
\beq
\label{self2}
\pa_-B=0.
\eeq
In the two--dimensional case the symmetry 3) becomes a global one,
just the shift by a constant, and the symmetry 2) assumes a slightly
different form. The action \eref{az2} is, in fact, invariant under 
the following two transformations:
\bea
\label{11}
1) & & \delta B = {\pa_-B\over \pa_-a} \varphi, \qquad 
\delta a = \varphi,\\
\label{22}
2) & & \delta B = \Lambda(a),\quad  \qquad \delta a = 0.
\eea
\eref{11} is just the transformation 1) of the previous section, written
in light--cone indices. In the transformation \eref{22} $\Lambda(a)$ is an 
arbitrary function of $a$, so this is not a local symmetry but 
rather an infinite set of global symmetries and at the quantum level it 
does not need a gauge--fixing. Nevertheless, it
is needed at the classical level
to obtain the self--duality relation \eref{self2}. To see this we
observe that \eref{emb} becomes in two dimensions (the equation for $a$
is again a consequence of this one)
$$
\varepsilon^{ij}\pa_i\left(v_jv_+\pa_-B\right)=0,
$$
whose general solution is $\pa_-B=\pa_- \tilde\Lambda(a)$. Performing a 
transformation 2), with $\Lambda(a)=\tilde\Lambda(a)$, we get $\pa_-B=0$.

We want now compute the effective action associated to \eref{az2}, in
an external gravitational field. Since only the symmetry 1) needs a gauge
fixing, formally this is given by
\beq
\label{formal}
e^{-\Gamma[e]}=\int {\cal{D}}B\, {\cal{D}} a\, e^{-S[B,a,e]}
\,\delta(a-a_0),
\eeq
where we introduced an arbitrary gauge--fixing function 
$a_0(x)$, and no Faddeev--Popov determinant arises. 

We evaluate  $\Gamma[e]$ in two steps. First we perform the functional
integral over $B$
\beq
\label{azeff}
e^{-\Gamma[e,a]}=\int {\cal{D}}B \, e^{-S[B,a,e]}.
\eeq
Since, as we will see, the symmetry 1) is anomaly free\footnote{The BRST
cohomology, associated to the symmetry 1), in the sector with ghost number
one - which is the one related to possible anomalies - is presumably
trivial.} 
one could expect 
that $\delta_1\Gamma[e,a]=\int d^2x{\delta\Gamma\over \delta a}\varphi=0$, 
which would imply that 
$\Gamma[e,a]$ is, actually, independent of $a$. But in a generic 
symmetry breaking regularization scheme\footnote{We will use a 
diffeomorphism preserving regularization which breaks local Lorentz 
transformations, Weyl transformations and the symmetry 1).}
one can have trivial anomalies, i.e. anomalies which have to be eliminated
by subtracting finite local counterterms. This will indeed be necessary in 
our case. After this subtraction we will perform the final 
$a$--integration, which will then become trivial as we will see. 

The main point is to show that $\Gamma[e]$ is, modulo local terms,
independent of $a_0$ and coincides with the effective action of a complex
Weyl fermion. To this end we recall now some known results regarding the 
determinantes and effective actions for two--dimensional non--chiral bosons 
and complex Weyl fermions.

The principal relation, for a generic metric $e_\pm^i$, is the following
\beq
\label{fund}
-{1\over 2}ln\, det\left(\sqrt{g}\bbox_g\right)=
ln \, det\left(\sqrt{g}\pa_-\right)+
ln \, det\left(\sqrt{g}\pa_+\right) + {\rm loc.}
\eeq
The l.h.s. is the effective action for a non chiral boson and at the
r.h.s. we have the sum of the effective actions for a left-handed
and a right-handed complex Weyl fermion. The local terms depend on
the regularizations. In a diffeomorphism preserving framework 
we have also the explicit expressions
\beq
\label{g+-}
\Gamma_\pm[e] \equiv  ln \, det \left(\sqrt{g}\pa_\pm\right)=
{1\over 96\pi}\int d^2x\sqrt{g}
\left(D_\mp\Omega_\pm {1\over \bbox_g} D_\mp\Omega_\pm \right),
\eeq
where
\bea
D_\pm&=&\pa_\pm \pm\Omega_\pm\\
\Omega_\pm &=&\pm {1\over \sqrt{g}}\pa_i\left(\sqrt{g}e_\pm^i\right).
\eea
In this framework the local terms in \eref{fund} are proportional
to $\int d^2x \sqrt{g}\,\Omega_+\Omega_-$. 

One more information we need is the anomaly carried by the chiral
determinants \eref{g+-} under {\it finite} Weyl rescalings
\beq
\Gamma_\pm[\lambda e]= \Gamma_\pm[e]+ {1\over 96\pi}
\int d^2x \sqrt{g}ln \lambda\left(\bbox_g ln\lambda 
\mp 2 ln\lambda D_\mp\Omega_\pm\right).
\eeq
The unique feature of this relation we will need is that under a finite 
rescaling of the metric $\Gamma_\pm[e]$ changes by terms which are local 
in the scaling parameter and in the metric itself.

The expected result for the effective action for chiral bosons is
\beq
\label{res}
\Gamma[e]=\Gamma_-[e],
\eeq
modulo local terms. 

To prove this result we proceed as follows. We begin by rewriting 
the action \eref{az2}, in terms of a fictitious metric, as
\beq
\label{azgf}
S[B,a,e]= {1\over 2}\int d^2x\sqrt{G}G^{ij}\pa_iB \pa_jB,
\eeq
where the metric 
$G^{ij}={1\over 2}\left(E_-^i E_+ ^j + E_+^i E_- ^j\right)$
is defined by 
\bea
\label{def-}
E^i_-&=&e^i_-\\
\label{def+}
E^i_+&=&e^i_+ -{\pa_+a\over \pa_-a}e_-^i
=-{2\over \pa_-a}\cdot{\varepsilon^{ij}\over \sqrt{g}}\pa_j a\\
\label{detm}
\sqrt{G}&=&\sqrt{g}.
\eea
The fact that the determinants of the two metrics coincide is a
consequence of the general relation \eref{det}. 
The intermediate effective action $\Gamma[e,a]$ can therefore be written in 
terms of the determinant of the laplacian associated to the fictitious metric
$G^{ij}$ 
\bea
\Gamma[e,a] &=& -{1\over 2} ln\, det \left(\sqrt{G}\bbox_G\right)\\
           &=& ln \, det\left(\sqrt{G}E_-^i\pa_i\right)+
            ln \, det\left(\sqrt{G}E_+^i\pa_i\right)\\ 
           &=& \Gamma_-[e]+ \Gamma_+[E].
\eea
Here we used the general decomposition \eref{fund}, applied to the metric
$G^{ij}$, and the relations \eref{def-},\eref{detm}. It remains to 
show that $\Gamma_+[E]$ is local in the fields $a,e$. 
This can be shown using the fact that for
a finite rescaling $\Gamma_+[E]$ changes by local terms 
$$
\Gamma_+[E]=\Gamma_+[E^*=\lambda E] + {\rm loc.}
$$
Choosing $\lambda={1\over\pa_-a}$ we get 
\bea
E^{*i}_+&=& {1\over \pa_-a} E^i_+ = -{2\over (\pa_-a)^2}\cdot
{\varepsilon^{ij}\over \sqrt{g}}\pa_j a\\
\sqrt{G^*}&=&(\pa_-a)^2\sqrt{G}=(\pa_-a)^2\sqrt{g}\\
\Omega_+^*&=&{1\over \sqrt{G^*}}\pa_i\left(\sqrt{G^*}E^{*i}_+\right)
={-2\over \sqrt{G^*}}\pa_i\left(\varepsilon^{ij}\pa_j a\right)=0.
\eea
Since $\Gamma_+[E^*]$ is quadratic in $\Omega_+^*$, see \eref{g+-}, 
it vanishes, actually. We have therefore 
$$
\Gamma[a,e]= \Gamma_-[e]+\Gamma_{\rm loc}[a,e].
$$
This proves, in particular, that the symmetry 1) is anomaly free. The
symmetry breaking term $\Gamma_{\rm loc}[a,e]$ is, in fact, local and
has to be subtracted, as mentioned above. After this subtraction
the integration over $a$ in \eref{formal} is now trivial since the 
integrand is $a$--independent, apart from $\delta(a-a_0)$, and the result
is $\Gamma[e]=\Gamma_-[e]$.

In practise, the functional integral in \eref{formal} can be evaluated
as
\beq
\label{start}
e^{-\Gamma[e]}=\int {\cal{D}}B \, e^{-S[B,a_0,e]},
\eeq
where finally one has to subtract all (local) terms which depend on
$a_0$.

This concludes the proof of \eref{res} at a non--perturbative level and
ensures, therefore, also that the gravitational anomaly carried by the 
chiral boson equals that of a complex Weyl fermion, which is associated
to the invariant polynomial ${1\over 96\pi}tr(RR)$, in agreement with
the index theorem for $D=2$ chiral bosons, \cite{AGG}.

Since in higher dimensions exact results for the effective actions
are not available 
it is also instructive to perform a perturbative derivation of 
$\Gamma[e]$. This will allow us to gain some insight in the 
ingredients which are essential for a perturbative evaluation of the 
contribution to the effective action which is responsible for
the anomaly also in higher dimensions. For the perturbative analysis
a convenient starting point is \eref{start}; so
we will also gain a better
understanding of how to deal with the non--manifestly 
covariant gauge fixings related with the particular choice of $a_0(x)$.

For the perturbative expansion, like we did in the preceding section,
we choose a class of gauge fixings parametrized by a constant vector 
$n_i$ satisfying
$$
n_in_j\eta^{ij}=1,
$$
and set
$$
a_0(x)=n_ix^i.
$$
In this case, since the effective action depends on $a_0$ only through
$\pa_ia_0=n_i$,  the term $\Gamma_{\rm loc}[a_0,e]$ becomes a local 
functional of only the metric, and $\Gamma[e]$ in \eref{start}
can depend on $n_i$ only through local terms.

We expand the metric around the flat one, 
$\eta^{ij}=\delta_-^{(i}\delta_+^{j)}$,
\bea
e_\pm^i&=&\delta_\pm^i+h_\pm^i\\
h_+^i&=&{1\over 2}\left(\delta_-^ih_{++}+ \delta_+^ih_{+-}\right)\\ 
h_-^i&=&{1\over 2}\left(\delta_+^ih_{--}+ \delta_-^ih_{-+}\right)\\
\sqrt{g}&=&1-{1\over 2}(h_{+-}+h_{-+})+o(h^2),
\eea
and expand the action in powers of $h_{\pm\pm}$ (from now on all
light--cone indices are flat, e.g. $n_\pm=\delta_\pm^i n_i,\, 
\pa_\pm=\delta_\pm^i\pa_i$)
\bea
\nonumber
S[B,a_0,e]=&-&{1\over 2}\int d^2xB\pa_-\left(\pa_+ -n_+^2 \pa_-\right)B\\
\nonumber
&+&{1\over 4}\int d^2x\left(\pa_+B-n_+^2\pa_-B\right)^2h_{--}\\
\label{esp}
&+& o(h^2).
\eea
In the first line we have the kinetic term and in the second the 
interaction term with the metric. Since we are only interested in the 
$h$-$h$ two--point function the higher order terms are not needed. We notice
that from the interaction terms $h_{-+}$ and $h_{+-}$
dropped out -- this is a 
consequence of Weyl invariance -- and that also $h_{++}$ decoupled.
This is due to the fact that left--handed chiral bosons,
are not coupled to this field in that its equation of motion
is $e_-^j\pa_jB=\pa_-B +h_-^j\pa_jB=0$. The action $S[B,a_0,e]$ itself
depends, actually, also on $h_{++}$, through the $o(h^2)$--terms,
but the effective action depends only locally on it.

From \eref{esp} we can read vertices and propagators
\bea
\label{prop}
B(k)-B(k)\qquad {\rm propagator} \qquad && {i\over k_-(k_+-n_+^2k_-)}\\
B(k)-h_{--}(p)-B(l)\qquad {\rm vertex}  \qquad &&
-{i\over 2}(k_+-n_+^2k_-)(l_+-n_+^2l_-)\equiv 
-{i\over 2}W(k)W(l).\nonumber\\
\label{vert}
\eea
In the computation of a one--loop Feynman diagram with $N$ external 
$h_{--}$ fields these vertices and propagators appear always in the sequence
$\cdots V\cdot P\cdot V \cdot P \cdots$, and from \eref{prop} and 
\eref{vert} one sees that 
the factor
$(k_+-n_+^2k_-)$ in the denominator of a propagator cancels always 
against a corresponding factor in an adjacent vertex. This means that
the effective propagator reduces simply to ${i\over k_-}$ which is the 
appropriate propagator for a chiral field in two dimensions. Moreover,
since the vertex is factorized, 
the above sequence can also be seen as a sequence of
building blocks of the form
\beq
\label{bb2}
{\cal{B}}(k)=
-{i\over 2}W(k){i\over k_-(k_+-n_+^2k_-)}W(k)
= {1\over 2} {k_+ (k_+-n_+^2k_-)\over k^2}={1\over 2}\left(
{k_+^2\over k^2}-n_+^2\right).
\eeq
We see in particular that in the $n$--dependent term of this basic
block the pole has cancelled. In the leading anomaly diagram, in the present
case the two--point function, this implies, for dimensional reasons, that
the $n$--dependence will occur only in local terms. For diagrams with 
more than two external $h_{--}$--legs the $n$--dependence will occur
also in non--local terms; but, since these terms have one or more
poles less, they will be cancelled by diagrams which originate from 
the $o(h^2)$ terms in \eref{esp}. For example, in the three--point function
$n$--dependent terms in which one pole cancelled lead to a contribution to the 
effective action of the form
$n_+^2\int d^2x\,d^2y\,h_{--}^2(x)G(x-y)h_{--}(y)$, and these cancel against 
contributions from the diagram 
with one vertex of the type \eref{vert} and one vertex of the type 
$B(x)h_{--}^2(x)B(x)$. 

The two--point function can now be easily evaluated. It is given by
a Feynman diagram with just two of the above building blocks \eref{bb2},
\beq
\Gamma_2(p)={1\over 8}\int{d^2k\over (2\pi)^2}\cdot
{k_+(k_+-n_+^2k_-) l_+(l_+-n_+^2l_-)
\over k^2l^2},
\eeq
where $l=p-k$, and can be evaluated with standard methods. A quick way to
do it is to change coordinates from $k_\pm$ to  $K_-=k_-$, $K_+=k_+-n_+^2k_-$,
and to define $P_-=p_-$, $P_+=p_+-n_+^2p_-$. This leads
to 
\beq
\Gamma_2(p)={1\over 8}\int{d^2K\over (2\pi)^2}\cdot{K_+^2(P-K)_+^2\over
K^2(K-P)^2},
\eeq
which gives, in dimensional regularization, 
$$
\Gamma_2(p)=-{1\over 192\pi}{P_+^3\over P_-}
=-{1\over 192\pi}\left({p_+^3\over p_-} -3p_+^2n_+^2+3n_+^4p^2-n_+^6p_-^2
\right).
$$
As anticipated above, the $n$--dependence is only in the local terms, 
which have to be subtracted,
and the non--local term amounts to a contribution to the effective action
given by
$$
\Gamma_2={1\over 4}\cdot {1\over 96\pi}\int d^2x\, \pa_+ \pa_+ h_{--}\,
{1\over \bbox}\, \pa_+ \pa_+ h_{--}.
$$
This coincides with the expansion of 
$
\Gamma_-[e],
$
see 
\eref{g+-}, up to local terms, since 
$$
D_+\Omega_-=-{1\over 2}\left(\pa_+ \pa_+ h_{--}-\bbox \,h_{+-}\right)
+o(h^2).
$$

This concludes the perturbative and non perturbative
analysis of the effective action for chiral bosons
based on the classical action \eref{az2}.
Many of the features appearing in the two--dimensional case will arise
also in higher dimensions, as we will see in the next section. One of the 
main points will be the determination of a convenient basic block, 
analogous to \eref{bb2}, and the determination of effective Feynman rules. 
We will also encounter a cancellation of factors between propagators
and vertices, as happened with the ones in \eref{prop} and \eref{vert}. 

The appearance of the common factor $(k_+-n_+^2k_-)$ is, actually, a 
consequence of the symmetry 2). Once one has chosen $a_0(x)=n_ix^i$,
the symmetry 2) reduces to a symmetry of the action $S[B,a_0,e]$. This 
action is now still invariant under $\delta B =\Lambda$,
but only for 
$\Lambda$'s  which depend on $x$ only through $n_ix^i$. The unique first order
derivative operator which is invariant under such transformations, due to 
$n_+n_-=1$,  is indeed
$(\pa_+- n_+^2\pa_-)B$, and this is the reason why it appears in 
\eref{esp} in the kinetic {\it and} interaction terms.

A similar role will be plaid in higher dimensions by the transformation
2) of the preceding section, which becomes then a true {\it local} symmetry. 
\vskip1truecm

\section{Chiral bosons in $4n+2$ dimensions}

In this section we want to show that the leading anomaly diagram in $D=2k+2$
($k$ even) dimensions, computed from our classical action \eref{az},
coincides with the diagram, based on conjectured Feynman rules, which has
been used by Alvarez--Gaum\'e and Witten
in \cite{AGW} to determine the anomaly. 
This is a one--loop diagram with $k+2$ external gravitons.

We outline first the procedure which has been adopted in \cite{AGW}
to conjecture these Feynman rules. One starts
from an action for non--chiral bosons in $D$ dimensions interacting with a
gravitational field. This is simply given by (as an integral over a 
$D$--form) 
\beq
\label{azAGW}
S[B,g]=-{1\over 2}\int H*H,
\eeq
where $H=dB$. First one derives the Feynman rules for non--chiral bosons
writing the metric as
\bea
g^{ij}&=&\eta^{ij}+h^{ij}\\
h&\equiv& \eta_{ij} h^{ij},
\eea
and expanding the action in powers of $h^{ij}$. For notational reasons it
is convenient to parametrize the symmetric matrix $h^{ij}$ in terms
of $D$ vectors $M_\alpha^i$, $\alpha=(1,\cdots,D)$ such that
$$
h^{ij}=\sum_\alpha M^i_\alpha M^j_\alpha. 
$$
This allows to introduce $D$ one--forms $M_\alpha=dx^iM_{\alpha i}$,
and we use the same notation $M_\alpha$ for the associated vectors
$M_\alpha^i\pa_i$, since no confusion could arise. The indices $i,j$ are now
raised and lowered with the flat metric and in what follows the sum 
over $\alpha$ will always be understood. This will allow us to 
write compact expressions for vertices and propagators.

With these notations 
the action \eref{azAGW} can be expanded as follows
\bea
S[B,g]&=&{1\over 2}\int \left(B*\bbox B + \delta B*\delta B\right)\\
        \label{vert1}
       &+&{1\over 2}\int dB\left(Mi_M-{1\over 2}h\right)*dB\\
       &+&o(h^2). 
\eea
The action \eref{azAGW} is invariant under the usual gauge 
transformations for $k$--form potentials; these can be fixed by adding to 
the action the "free" i.e. metric--independent term 
$-{1\over 2}\int \delta B*\delta B$, and the propagator becomes then
simply 
\beq
\label{propA}
{\rm propagator}\qquad -{1\over \bbox}\quad .
\eeq
For non--chiral bosons the $BhB$--vertex could be read from \eref{vert1}.
For chiral bosons the authors of \cite{AGW} conjectured Feynman rules
for which the propagator is still given by \eref{propA} while, for 
what concerns the
vertex, they inserted in \eref{vert1} on $dB$ the projector ${1\over 2}
(1+*)$, i.e. they took as interaction term, instead of
\eref{vert1} the expression
\beq
\label{vert2}
{1\over 2}\int 
{(1+*)\over 2}\,\,dB\left(Mi_M-{1\over 2}h\right){(1+*)\over 2}\,\,dB.
\eeq
From this expression one can read the $B(k)$-$h(p)$-$B(l)$ vertex which,
schematically, is given by an expression of the form (it is convenient
to keep the external leg $h^{ij}$ inserted)
$$
k^i\left(Mi_M-{1\over 2}h\right)l^j,
$$
which is a $k\times k$ antisymmetric tensor. In a one--loop diagram 
the sequence $\cdots V\cdot P\cdot V \cdot P\cdots$ can then be written
as
$$
\cdots k^i\left(Mi_M-{1\over 2}h\right)l^j\cdot {1\over l^2}\cdot
l^r\left(Mi_M-{1\over 2}h\right)q^s \cdots .
$$
From this one can extract a building block which depends only on a single 
momentum, say $l$
$$
\left(Mi_M-{1\over 2}h\right)l^j\cdot {1\over l^2}\cdot
l^r.
$$
This is now a $(k+1)\times (k+1)$ antisymmetric tensor, so, turning to
configuration space, it can be represented as a linear operator
which sends a $(k+1)$--form in a $(k+1)$--form. Reinserting the appropriate
contraction of indices this operator is given by
\beq
\label{blockAGW}
{\cal{B}}_{AGW}=-\left(Mi_M-{1\over 2}h\right){1\over 2}(1+*)\,\, 
{d*d\over \bbox}\,\,{1\over 2}(1-*). 
\eeq
Since we have the operatorial identity 
\beq 
\label{ident}
{1\over 2}(1+*) \left(Mi_M-{1\over 2}h\right){1\over 2}(1+*) =0,
\eeq
and the one--loop diagram is now a chain of blocks \eref{blockAGW}, the last 
projector in the block can be omitted. 
 
In the remaining part of this section we want now show that the action 
\eref{az} leads to the same building block ${\cal{B}}_{AGW}$.

Starting from this action, the effective action $\Gamma[g]$ 
is obtained via a functional
integral over $B$ and $a$ upon gauge fixing the local symmetries 1)--3). 
For the symmetry 1) we proceed as in the perturbative treatment 
of the two--dimensional case, inserting 
the $\delta$--function $\delta(a-n_ix^i)$. Since also 
in higher dimensions the symmetry 1) is 
expected to be anomaly free, the effective action will depend on $n$ 
only through local terms. 

For what concerns the symmetries
2) and 3), we use the gauge--fixings $2')$ and $3')$ of section two,
but now with an appropriate weighting function $f(b_2,b_3)$:
\bea
\nonumber
e^{-\Gamma[g]}&=&
\int {\cal{D}}B\,{\cal{D}}a\,e^{-S[B,a,g]}
 \int{\cal{D}}b_2\, {\cal{D}}b_3 \,\delta(a-n_ix^i)\delta(i_nB-b_2) 
\delta(\delta B-b_3)\,
e^{-{1\over 2}\int f(b_2,b_3)}\\
\label{AzeffD}
&=&\int {\cal{D}}B\, e^{-S_n[B,g]},
\eea
where
\beq
\label{azgfD}
S_n[B,g]=S[B,n_ix^i,g]+{1\over 2}\int f(i_nB,\delta B).
\eeq
Here $b_2$ and $b_3$ are $(k-1)$--forms, $f(b_2,b_3)$ is a quadratic 
metric--independent function of these fields which parametrizes the
gauge fixing terms (contractions are made with the flat metric), and in the
$\delta$--functions appearing in \eref{AzeffD} the terms $i_n B$ and
$\delta B$ are also constructed with the flat metric. This implies that the
gauge fixing term in \eref{azgfD}, ${1\over 2}\int f(i_nB,\delta B)$,
is metric independent and that no Faddeev--Popov determinants arise.
For the moment $f$ is left undetermined, we will make a
convenient choice below. 

The gauge--fixed action $S_n[B,g]$ can now be expanded in powers of $h^{ij}$ 
and, using the same notation as above, one gets
\bea
\label{kin3}
S_n[B,g]=&-&{1\over 2}\int\left(B*\left[T\pa_n+T^2\right]B 
-f(i_nB,\delta B)\right)\\
\label{vert3}
&+&{1\over 2}\int TB *\left[Mi_M+{1\over 2}h-(n\cdot M)^2 +(n\cdot M)
*Mn\right]TB\\
&+&o(h^2).
\eea
Here with $(n\cdot M_\alpha)$ we mean $n_iM_\alpha^i$ and the operator
$T$ has been defined in section two, \eref{T}. It plays the same role
as the differential operator $(\pa_+- n_+^2\pa_-)$ in the two--dimensional
case; for $D=2$ it reduces, actually, apart from a constant, to this 
operator. Once one has chosen $a_0=n_ix^i$, the symmetry 2) reduces, indeed, 
to $\delta B=n \Lambda_{k-1}$ and it is precisely the combination 
$TB=*ndB$ which is invariant under this reduced local symmetry and under the 
usual gauge transformation 3).

We choose now the function $f$  such that the kinetic operator in
\eref{kin3} becomes as simple as possible. The explicit expression
of the operator $T^2$ is given in \eref{t2} and it can be reduced
simply to $\pa_n^2-\bbox$ upon choosing
$$
f(b_2,b_3)=db_2*db_2 +b_3*b_3 +2b_2*\pa_n b_3.
$$
With this choice one obtains for the kinetic term
\beq
-{1\over 2}\int \left(B*\left[T\pa_n+T^2\right]B -f(i_nB,\delta B)\right)
=-{1\over 2}\int B*\Omega B,
\eeq
where the gauge--fixed kinetic operator is
\beq
\Omega=\pa_n^2-\bbox +\pa_nT.
\eeq
It sends a $k$--form in a $k$--form and becomes, in momentum space, 
a $k\times k$ antisymmetric tensor. The $B$-$B$ propagator $P$ is just the 
inverse and can be easily  computed using algebraic
methods. The essential ingredient is the identity \eref{t3} which implies
that every power series in  $T$, like $\Omega^{-1}$, 
can be reduced (in momentum space) to a polynomial in ${\msbm I}$, $T$ and 
$T^2$. In configuration space the result is 
\beq
P=\Omega^{-1}={1\over \pa_n^2-\bbox}
\left({\msbm I}+{T\pa_n\over \bbox}-{T^2\pa_n^2\over\bbox\,(\pa_n^2-\bbox)}
\right).
\eeq
Since in the interaction term \eref{vert3} $B$ appears always as $TB$, what is 
actually needed in a Feynman diagram is the combination
\beq
\label{comb}
TPT=\left(\pa_n-T\right)T\,{1\over \bbox},
\eeq
which leads finally to the cancellation of the unphysical pole 
${1\over \pa_n^2-\bbox}$ and to the appearance of the massless
physical pole ${1\over \bbox}$.

Now we step to the problem of individuating a convenient building block
for a one--loop Feynman diagram with a certain number of external 
gravitons, the one responsible for the gravitational anomaly carrying 
$k+2$ of them.

If we indicate the operator between square brackets in \eref{vert3} with 
$W\equiv W(h)$, it sends a $k$--form in a $k$--form, 
the interaction term can be written as
\beq
{1\over 2}\int TB * W TB,
\eeq
and the vertex--propagator sequence becomes, due to \eref{comb},
\beq
\cdots \left[TWT\right]P\left[TWT\right]\cdots
=\cdots TW\left(\pa_n-T\right)T\,{1\over \bbox}\,WT\cdots
\eeq
We can extract as building block
$$
-W\left(T-\pa_n\right)T\,{1\over \bbox}=
-W\left(T-\pa_n\right)*d\,{1\over \bbox}\,n,
$$
or, equivalently, due to cyclicity
\bea
{\cal {B}}&=&  -n\,W\left(T-\pa_n\right)*d\,{1\over \bbox}\\
&=& -n\,W\left(T-i_nd-di_n\right)*d\,{1\over \bbox}\\
\label{drops}
&=& -n\,W\left(T-i_nd\right)*d\,{1\over \bbox}
-n\,\left[Wd*d\,{1\over \bbox}\right]n.
\label{last}
\eea
We used the identity $\pa_n=i_nd+di_n$ in the first line and \eref{v2}
to get the last line. 

In the sequel we will make repeated use of the 
identities \eref{v1}-\eref{v4} with $v=n$, $v^2=1$. 

This block is written as an operator which sends a $(k+1)$--form in a 
$(k+1)$--form, as is realized by direct inspection, so in momentum space
it becomes a $(k+1)\times (k+1)$ antisymmetric tensor, as does 
${\cal {B}}_{AGW}$. 
In a one--loop diagram these blocks are multiplied by themselves; the  
term between square brackets in the last line carries a factor of $n$ on 
each side, so one of these factors encounters necessarily another factor of 
$n$ and, due to antisymmetry (or, due to the fact that the square of a 
one--form is zero), these terms drop all out. We remain therefore only with 
the first term in \eref{drops}. Inserting the definition of $T$,
this can be written as
$$
{\cal {B}}=n\,W\,i_n(1+*)\,{d*d\over \bbox}.
$$
Due to the appearance of the projector $(1+*)$, one can now eliminate the
unique $*$-operator contained in $W$. After some algebra one finds
\beq
\label{form}
{\cal {B}}=-n\,i_n\left(Mi_M-{1\over 2}h\right)(1+*)\,{d*d\over \bbox}.
\eeq 

The appearance of the combination $\left(Mi_M-{1\over 2}h\right)$ in this
formula, as well as in \eref{blockAGW}, is due 
to Weyl--invariance at the linearized level, i.e. invariance under
$\delta h^{ij}(x)=\lambda(x)\eta^{ij}$.

In \eref{form}, due to the identity \eref{ident}, one can insert the 
projector ${1\over 2}(1-*)$  after the operator $n\,i_n$ and, due to 
cyclicity, one can replace \eref{form} with 
\bea
{\cal {B}}&=&-\left(Mi_M-{1\over 2}h\right)(1+*)\,\,{d*d\over \bbox}\,\,
ni_n\,{1\over 2}(1-*)\\
&=&-\left(Mi_M-{1\over 2}h\right)(1+*)\left({d*d\over \bbox}\,{1\over 2}
(1-*) +{1\over 2}\right)ni_n\,{1\over 2}(1-*)\\
&=&-
\label{finale}
\left(Mi_M-{1\over 2}h\right){1\over 2}(1+*)\,\,{d*d\over \bbox}\,\,{1\over 2}
(1-*)
-\left(Mi_M-{1\over 2}h\right){1\over 2}(1+*)\,ni_n\,{1\over 2}(1-*).
\nonumber
\\
\eea
In the second line we used \eref{v4}, for a flat metric, and in the third
the identity \eref{v33} with $p=k+1$, which is odd.
  
This is the formula for the building block which generalizes \eref{bb2}
to a generic dimension, for  $D=2$ it reduces actually to that formula. 

Again we see that in the $n$--dependent term the pole (propagator)
${1\over \bbox}$
cancelled out, so, as in the two--dimensional case, the  $n$--dependence
in these terms has to cancel against diagrams which contain vertices of
the type $B(h^{ij})^pB$. For what concerns the leading part of the anomaly,
these diagrams have one (or more) propagators less and they can not 
contribute to the anomaly \cite{BN2}, \cite{AGW}.

The first term in \eref{finale} coincides with \eref{blockAGW}, and hence the
gravitational anomalies derived from the classical action \eref{az}
coincide with the ones computed in \cite{AGW}.
\vskip1truecm

\section{Final remarks}

In this paper we proved that the gravitational anomalies derived from 
the classical manifestly invariant action for chiral bosons in
$4n+2$ dimensions, proposed in \cite{PST}, coincide with the expected
ones. This supports the quantum reliability of the new method itself
at the perturbative level. On the other hand, as all lagrangian formulations 
of theories with chiral bosons, the method is expected to be insufficient
for what concerns the quantization of these actions on manifolds 
with non trivial topology \cite{Witten}; see, however, also 
\cite{Dolan}.

In this paper we were concerned with diffeomorphism anomalies of 
the ABBJ--type, 
which are non trivial cocycles of the corresponding BRST operator,
and exist only in $4n+2$ dimensions (clearly they can be shifted to
Lorentz--anomalies). In a generic even dimension,
however, there exists also another class of diffeomorphism 
cocycles\footnote{None of these cocycles contains the $\varepsilon$--tensor.},
of the ``Weyl--type'', 
which can be eliminated at the expense of Weyl--anomalies \cite{Math},
if the corresponding theory is classically Weyl invariant (otherwise
they become simply trivial diffeomorphism cocycles). 
The resulting inequivalent Weyl--cocycles, in four and six dimensions,
have been determined through a cohomological analysis
in \cite{Weyl}, see also \cite{DS,Duff}. In four dimensions there are
three of them and in six dimensions there are four. 

Our classical action for chiral bosons \eref{az} is indeed invariant
under local Weyl rescalings, $ g^{ij}\rightarrow e^\lambda g^{ij}$,
$B\rightarrow B$, $a\rightarrow a$; therefore one expects that, as in 
the two--dimensional case, the effective action $\Gamma[g]$ is plagued
also by diffeomorphism anomalies of the Weyl--type, or, equivalently,
by the Weyl--anomalies discussed in \cite{Weyl}. Having at our
disposal a classically manifestly invariant 
action principle for chiral bosons in an external gravitational
field could be essential in the determination of these anomalies, which, 
to our knowledge, for $D> 2$ are still unknown. In six 
dimensions, for example, this would amount to the determination
of the coefficients of the four non trivial Weyl--cocycles mentioned above.
In higher dimensions, even the form (and the number) of the non trivial
cocycles is unknown.

The fact that the action \eref{az} gave rise to the correct ABBJ
gravitational anomalies carried by chiral bosons,
makes us hope that it can also prove useful
to make some progress for what concerns  the determination of their
Weyl anomalies.

\paragraph{Acknowledgements.}
\ We are grateful to P. Pasti and M. Tonin for their interest 
in this work and useful discussions. This work was supported by the 
European Commission TMR programme ERBFMPX-CT96-0045.

\vskip1truecm

\end{document}